# Prioritising Server Side Reachability via Inter-process Concolic Testing

Maarten Vandercammen[a], Laurent Christophe[a], Dario Di Nucci[b], Wolfgang De Meuter[a], and Coen De Roover[a]

a   Vrije Universiteit Brussel, Brussels, Belgium
b   Jheronimus Academy of Data Science, 's-Hertogenbosch, The Netherlands

**Abstract**   **Context:** Most approaches to automated white-box testing consider the client side and the server side of a web application in isolation from each other. Such testers lack a whole-program perspective on the web application under test.
   **Inquiry:** We hypothesise that an additional whole-program perspective would enable the tester to discover which server side errors can be triggered by an actual end user accessing the application through the client, and which ones can only be triggered in hypothetical scenarios.
   **Approach:**   In this paper, we explore the idea of employing such a whole-program perspective in *inter-process* testing. To this end, we develop STACKFUL, a novel concolic tester which operates on full-stack JavaScript web applications, where both the client and the server side are JavaScript processes communicating via asynchronous messages — as enabled by e. g., the WebSocket or Socket.IO-libraries.
   **Knowledge:**   We find that the whole-program perspective enables discerning *high-priority* errors, which are reachable from a particular client, from *low-priority* errors, which are not accessible through the tested client. Another benefit of the perspective is that it allows the automated tester to construct practical, step-by-step scenarios for triggering server side errors from the end user's perspective.
   **Grounding:**   We apply STACKFUL on a collection of web applications to evaluate how effective inter-process testing is in distinguishing between high- and low-priority errors. The results show that STACKFUL correctly classifies the majority of server errors.
   **Importance:**   This paper demonstrates the feasibility of inter-process testing as a novel approach for automatically testing web applications. Classifying errors as being of high or low importance aids developers in prioritising bugs that might be encountered by users, and postponing the diagnosis of bugs that are less easily reached.

# The Art, Science, and Engineering of Programming

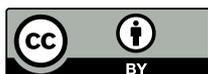

**Prioritising Server Side Reachability via Inter-process Concolic Testing**

## 1 Introduction

Since the advent of Node.js,[1] JavaScript is an increasingly popular choice for implementing both the client and the server side of highly interactive web applications such as collaborative editors, chat boxes, and file sharing services. These applications are often referred to as *full-stack* web applications, as their entire technology stack stems from the JavaScript ecosystem. This stands in contrast to more traditional web applications where the server side might employ e.g., PHP or Java. The dynamic nature of JavaScript, which features prototype-based inheritance, dynamic code evaluation, and dynamic property creation and deletion, renders static verification of JavaScript programs hard [16, 22]. As a result, several automated testing tools have become available for web applications [19, 32, 33] These tools typically consider the client side and the server side of the application under test in isolation from each other, i.e., they employ an *intra-process* approach towards testing.

In this paper, we demonstrate how a whole-program perspective on the web application enables the tester to distinguish various levels of *importance* for the bugs it reports. Whether or not a bug on the server or the client side of a web application may manifest itself in real-life scenarios depends on how these sides interact with each other. We argue that, for any given configuration of a client and a server, bugs that can arise in that configuration are of higher importance than bugs that can only arise in hypothetical, untested configurations. Although both types of bugs should be diagnosed and corrected, developers may opt to prioritise the former.

We present a two-phase approach to detecting and distinguishing the importance of the bugs in a full-stack web application. An intra-process phase tests the server side in isolation first, which is followed by an inter-process phase in which the tester explores the entire application under a given client-server configuration and attempts to exercise the client in such a way as to reach the previously discovered server errors. This second phase enables the tester to discern *high-priority* errors that may arise under the given configuration, from *low-priority* errors that cannot—although these errors may still arise under a different, untested configuration.

The ability to discern false positive from true positive errors has been shown to increase developers' confidence in bug detectors [18, 23, 25]. We argue that the ability to distinguish high and low priority errors has comparable effects on their confidence in dynamic testers. Moreover, having the tester exercise a particular client-server configuration facilitates generating practical, step-by-step bug reports that precisely describe how an end-user might stumble upon a bug in a real-life setting. Taking a whole-program perspective becomes even more compelling when considering the increasingly full-stack nature of web applications. This evolution makes it feasible for a single tool to simultaneously and uniformly test both the client and the server processes.

---
[1] https://nodejs.org/en/, last accessed 2020-10-01.

Our work makes the following contributions:
- We illustrate the potential benefits of automated inter-process testing of full-stack web applications using a motivating example.
- We present the design and prototype implementation of STACKFUL, the first concolic tester that automatically explores full-stack JavaScript web applications in an inter-process manner.
- We compare the inter-process and intra-process approaches to concolic testing empirically on their ability to discern bugs that can arise in a given client-server configuration from those that cannot.

The remainder of this document is structured as follows: section 2 introduces concolic testing of regular and of event-driven applications as the background for this work. Section 3 uses a motivating example to illustrate how intra-process automated testing may lose accuracy on full-stack programs. Section 4 presents the design and implementation of STACKFUL, an inter-process concolic tester for full-stack web applications. Section 5 presents a formal description that outlines the intra- and inter-process testing phases in more detail. We evaluate STACKFUL and compare the intra-process and inter-process approaches to concolic testing in section 6, and discuss these results in section 7. Finally, section 8 presents the related work, and section 9 concludes the paper and proposes avenues for future research.

## 2 Background

We introduce concolic testing and automated testing of event-driven applications as the background for our work.

### 2.1 Concolic Testing

*Concolic testing* [15, 31] is an automated approach to white-box testing. The goal of a concolic tester is to explore all feasible program paths by iteratively manipulating the values of non-deterministic variables in the program such as user and file input, and random numbers [2, 6]. To this end, the concolic tester performs *concrete* and *symbolic* execution simultaneously. Indeed, on the one hand, the tester steers the *concrete* execution of the program along an extension of a previously-explored path and reports newly-encountered errors. On the other hand, the *symbolic* execution gathers existing and newly-encountered non-deterministic variables that constrain the program's execution so that subsequent test runs will explore different program paths.

We illustrate the inner-workings of a concolic tester via the example in JavaScript depicted in figure 1. Lines 13 and 14 assign random values to the variables x and y. As sources of non-determinism among the program's executions, the tester represents them symbolically as $x$ and $y$. Suppose that in the first test run, the concolic tester randomly assigns the values 3 to x and 5 to y. These values cause the condition on line 7 to be false, and the program terminates without errors. Simultaneously to

```
1  function twice(v){
2    return v * 2;
3  }
4
5  function f(x, y){
6    var z = twice(y);
7    if(z==x) {
8      if(x > y + 10) {
9        throw new Error();
10     }
11   }
12 }
13 var x = Math.random();
14 var y = Math.random();
15 f(x,y);
```

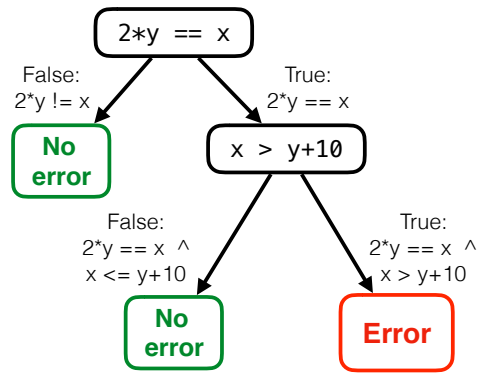

**Figure 1** A JavaScript program and its corresponding symbolic execution tree, after a translation to JavaScript of figure 1 in [5].

this concrete execution, the tester also collects the symbolic representation of the conditional predicate that was encountered on line 7 in the form of a so-called *path constraint*, i.e., $2y \neq x$. After completing this run, the tester attempts to explore another path, such as the path leading to the if-statement on line 8. The tester negates the path constraint to this end and feeds the resulting $2y = x$ to an SMT solver, which finds a solution that assigns e.g., 2 to $x$ and 1 to $y$. The concolic tester re-executes the program and assigns the value 2 to x and 1 to y. Concrete execution reaches the if-statement on line 8, then takes the else-branch there and the program terminates again without errors. In the meantime, the symbolic execution gathered the path constraint $2y = x \land x \leq y + 10$. The tester negates the last element of this path constraint and feeds the resulting path constraint $2y = x \land x > y + 10$ into the SMT solver, which finds e.g., the values 30 and 15 for x and y as a solution. A new test run is started with these values, and the concrete execution reaches the error on line 9, which is reported by the tester. As no new branches were encountered by the tester during this last run, the tester deduces that it has explored all feasible program paths and terminates. In practice, for realistic programs with a (near-)infinite number of program paths, the testing phase is terminated either upon exceeding a given time limit or test budget or when the desired level of code coverage has been reached. The three path constraints that were found can be collected in a symbolic execution tree that represents all possible executions of the program, as shown in figure 1.

## 2.2 Concolic Testing of Event-Driven Applications

In the case of event-driven applications, any automated tester that aims to systematically explore all possible program paths must consider not only several conditional branches but also several sequences of events during the program's execution. Consider the code snippet in listing 1.

■ **Listing 1** An event-driven JavaScript program

```
1  var counter = 0;
2  button1.addEventHandler("click", function (e) {
3    if (generateRandomInt() % 2 === 0) {
4      console.log(counter);
5    } else {
6      counter = 0;
7    }
8  }
9  button2.addEventHandler("click", function (e) {
10   counter++;
11 }
```

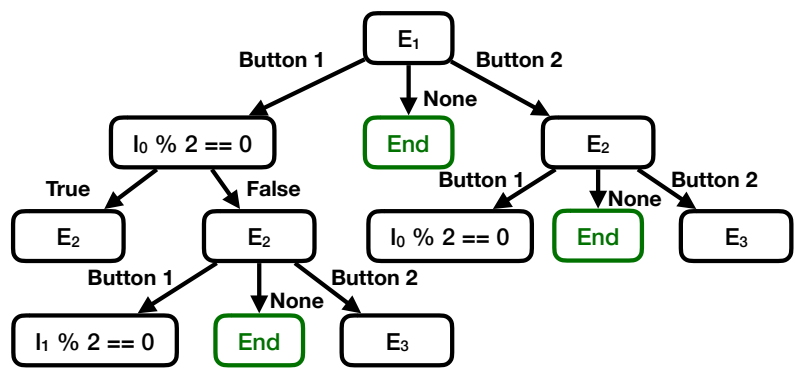

■ **Figure 2** Part of the (infinite) symbolic tree produced by systematically testing the code of listing 1

The exact behaviour of this code snippet depends on the order in which and the number of times both buttons are clicked, as well as the value of the random number that is generated each time button1 is clicked. In fact, apart from the registration of these event handlers, no code is executed at all *until* a click event is triggered for either button. A naive, but complete, exploration of all possible program paths must therefore consider both the values of symbolic input parameters, such as the randomly generated integer at line 3, as well as the sequence of events to be followed. Part of the symbolic execution tree representing all possible program paths for this code snippet is depicted in figure 2.

The root node of this symbolic tree corresponds to the *event* $E_1$ that is triggered first. As the program defines two event handlers, there are three possible outcomes: (i) button1 is clicked, (ii) button2 is clicked, or (iii) *no* button is clicked at all. In the last case, the program terminates. If button1 is clicked, the execution splits again into two possible paths: (i) one where the expression generateRandomInt() % 2 === 0 is true and (ii) one where it is not. Whatever button is clicked, once the execution of the corresponding event handler completes, the previous three-fold choice is presented again for a potential second event $E_2$.

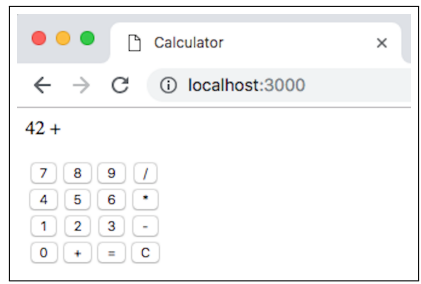

**Figure 3** The CALCULATOR application

## 3 Motivating Example

We illustrate the benefits that a whole-program perspective can bring to concolic testing of full-stack web applications through CALCULATOR. Figure 3 depicts the client side of this application as rendered by a browser. The user may enter arithmetic expressions in the form *n1 op n2*, where *n1* and *n2* are numbers, and *op* is an arithmetic operator. When the user presses the button labelled =, the expression is sent to the server where its result is calculated. Then, this result is sent back to the client to be displayed.

Listing 2 depicts an extract from the client's implementation. The client calls function io (line 1) for a connection to the server through a bidirectional socket. Afterwards, it registers separate event handlers for a mouse click on each individual button. Note that the event handlers for most buttons have been elided from the listing. Importantly, the event handler for the button labelled = calls the compute function (line 4). The client represents the arithmetic expression as an object input (line 6) containing three fields left, op, and right. Function compute checks whether the expression that was entered is a valid arithmetic expression (line 8). If the check fails, the function shows an appropriate error message to the user (line 9). Otherwise, the function sends the input to the server through the socket (line 11). Finally, the client registers a callback for messages from the server (line 14). After the server completes the computation, it sends the result to the client through the result parameter of the callback. Upon receiving such a message, the client shows the result to the user (line 16).

Listing 3 depicts an extract of CALCULATOR's server side. The code creates an http-server instance (line 2) that listens to incoming connections from clients (line 3). When a new client connects, the corresponding callback is triggered (lines 3–24) with the socket through which the client is connected as an argument. The server registers a callback (line 5) on each socket to listen for compute messages coming from the corresponding client. When the client sends such a message, the server retrieves the left and right operand, as well as the operator (lines 7–9). The result is computed from these three elements (lines 11–20) and sent back to the client via its socket (line 22). Importantly, the server throws an error when it detects a division by zero (line 16) or when it does not recognise the operator to be applied (line 19).

**▪ Listing 2** Part of the client side code for the CALCULATOR program

```
1  var socket = io(); // Connect with the server
2  document.getElementById("0").addEventListener("click", (evt) => clickDigit(0));
3  document.getElementById("+").addEventListener("click", (evt) => clickOperator("+"));
4  document.getElementById("=").addEventListener("click", (evt) => compute());
5  … // Register event handlers for other buttons
6  var input = {left: 0, op: "", right: 0}; // Arithmetic exp
7  function compute() {
8    if (! isValidExpression(input)) {
9      resultElement.innerHTML = "Expression is invalid";
10   } else {
11     socket.emit("compute", input); // Send the expression to the server
12   }
13 }
14 socket.on("result", function (result) {
15   // Receive computation result from server
16   resultElement.innerHTML = result; // Show the result
17 });
```

**▪ Listing 3** Part of the server side code for the CALCULATOR program

```
1  … // Setting up the server
2  var io = require('socket.io')(…);
3  io.on("connection", function(socket) {
4    // A new client has connected
5    socket.on("compute", function (input) {
6      // Receive input from client
7      var left = input.left;
8      var right = input.right;
9      var op = input.op;
10     var result;
11     switch (op) {
12       case "+": result = left + right; break;
13       case "-": result = left - right; break;
14       case "*": result = left * right; break;
15       case "/": result = if (right === 0) {
16         throw new Error("Dividing by zero");
17       }
18       result = left / right; break;
19       default: throw new Error("Unknown operator");
20     }
21     // Send the result to the client
22     socket.emit("result", result);
23   }
24 });
```

### 3.1 Testing CALCULATOR

Traditional concolic testing would test both sides of the application in isolation from each other. When applied to the client, such a tester should exercise the event handlers of all buttons and the callback for receiving the server message containing the result of the computation (lines 14–17). This callback could be exercised either by mocking the server and generating messages containing random result values, or by actually requiring the testing set-up to run a server besides the client process under test, in which case the client may send concrete messages to the server and expect real messages back. A traditional concolic tester should be able to achieve 100 % line coverage for the client in either case.

When testing the server, the tester could again opt to exercise the computation request callback (lines 5–24) by mocking client messages. However, as the server is being tested in isolation from the client, the tester does not have any information on the contents of the message and can therefore only assume that the message may contain any operand and operator. In practice, when combining this server code with the previously described client code, it is clear that the error at line 19 is unreachable, as this particular client ensures that only messages using any of the four valid operators are sent to the server. On the other hand, the division-by-zero error at line 16 is still reachable on the server, as this client does not check for this error. We therefore say that the division-by-zero error is a *high-priority* error, while the invalid operator error is a *low-priority* error. In the case of intra-process testing, the tester only considers one of either the client or the server processes. To test these processes, it must mock user-triggered events and inter-process messages. In the case of the latter though, no constraint can be placed on these messages as the tester does not know where these messages would come from nor how they would be produced. It is therefore unable to distinguish between the importance of both errors.

In the case of inter-process testing, the tester considers both the client and the server processes. Messages that are sent by this particular client as a result of user-triggered events can be tracked as they reach the server, making the tester aware of *how* these messages were produced and which input values for these messages are likely. In the case of this CALCULATOR application, messages with an invalid operator would never be sent to the server but would be halted by the client at line 8. When combined with traditional, intra-process testing of the server in isolation, the tester can realise that the division-by-zero is a high-priority error, as it can be triggered by any user exercising this client. The "Unknown operator" error is a low-priority error as it can only be triggered in combination with a different client, or by circumventing the client altogether. This issue is visualised for a generic full-stack application in figure 4 and 5.

### 3.2 Actionability of the Bug Report

Another benefit of the whole-program perspective taken by inter-process testers, is that it enables generating bug reports that more closely resemble how a user might encounter the bug in a real-world setting. Apart from printing the stack trace and the location in the code where these errors occurred, intra-process testers could also

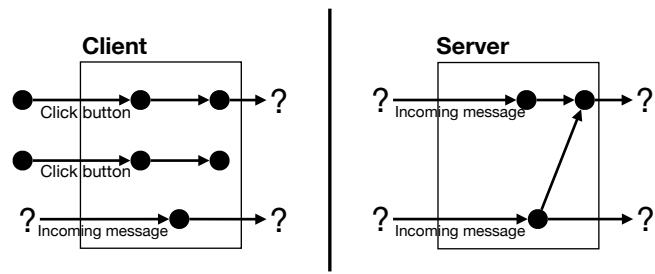

**Figure 4** Intra-process testing of a client and a server in isolation from each other

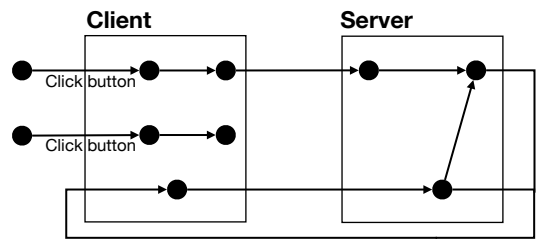

**Figure 5** Inter-process testing of a particular client-server configuration

```
(Server): Tester detected error in file "index.js", at position (22:4)
ERROR: Dividing by zero
Error encountered by triggering the following user events:
Clicked button "Button1"
Clicked button "Button/"
Clicked button "Button0"
Clicked button "Button="
```

**Figure 6** Example of a bug report specifying how an error on the server can be encountered by triggering user events on the client

specify that these errors were triggered because the server received a compute message with the appropriate values for the operator and operands. However, inter-process testers can go one step further: because they analyse both sides of the application simultaneously, it is trivial for them to rephrase this report in terms of which buttons must be clicked on the client side to trigger the errors on the server side. Such a rephrasing might be useful for very complex applications, where it is not obvious *how* a faulty message can be sent in practice. In those cases, it might be convenient for the developer to have a practical step-by-step report to reproduce the error available. An example of such a report, stipulating how the division-by-zero error on the server can be triggered, is depicted in figure 6.

## 4 STACKFUL: An Inter-process Tester for Web Applications

To analyse the differences between intra-process and inter-process concolic testing, we have developed STACKFUL, a tool capable of employing both variants. STACKFUL focuses on automated testing of web applications where both the client and the server are written in JavaScript, i.e., full-stack JavaScript applications, and where these tiers communicate via WEBSOCKET or SOCKET.IO messages. These libraries are a popular choice for implementing *bidirectional* communication between a client and the server, which is much more cumbersome when using HTTP requests alone.

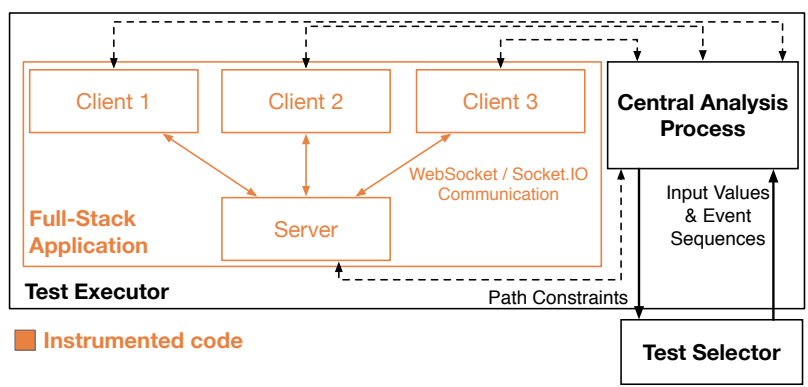

**Figure 7** The architecture of STACKFUL.

STACKFUL operates in two phases. In the first phase STACKFUL automatically tests the server side of the web application in isolation, i.e., it performs intra-process testing of the server component. In the second phase, STACKFUL performs inter-process testing of a particular client-server configuration to attempt to exercise the client in such a way that the previously discovered server errors are triggered. Once testing terminates, any error that was reproduced via this particular client-server configuration is labelled a *high-priority* error, while errors that could not be reproduced are labelled *low-priority*. In this section, we provide an overview of STACKFUL and highlight some of our design decisions.

In both the first and the second phase, STACKFUL leverages ARAN-REMOTE, a dynamic analysis platform [10] for distributed JavaScript applications. ARAN-REMOTE analyses are executed in a centralised process, called the *central analysis process*, which communicates with the processes under analysis. Under the hood, ARAN-REMOTE relies on ARAN [9], a state-of-the-art JavaScript program instrumenter to deploy the dynamic analyses.

Figure 7 depicts STACKFUL's components, as well as the processes, shown in orange, of the full-stack application under test. These processes consist of one server process, running in NODE.JS, and one or more client processes, each running in a browser. Both kinds of processes are instrumented by ARAN. STACKFUL itself consists of:

- The *central analysis process*, executing on ARAN-REMOTE, which is responsible for: (i) gathering intra-process path constraints, (ii) building an inter-process overview, (iii) generating program input values, and (iv) mocking user-events and messages to execute their associated handlers.
- A *test selector* which maintains the symbolic execution tree and suggests new program paths to explore. To this end, it must also suggest appropriate event sequences.

■ **Listing 4** An example of the instrumentation generated by Aran

```
1  if (_.test(_.binary("===", _.read($right, "right"), _.primitive(0)))) {
2    throw _.throw(_.construct(global.Error, [_.primitive("Dividing by zero")]));
3  }
4  $result = _.write(_.binary("/", _.read($left, "left"), _.read($right, "right")), "result");
```

In the remainder of this section, we first discuss how StackFul instruments an application to simultaneously apply symbolic and concrete execution (section 4.1) and how the central analysis process and the test selector collaborate to test a program in an intra-process manner (section 4.2 and 4.3). Section 4.4 describes the second phase of StackFul, in which errors discovered during the intra-process phase are reproduced. Section 4.5 summarises the challenges in both phases that must be overcome by StackFul.

### 4.1 Collecting Path Constraints

In Aran-Remote, an analysis consists of a set of functions, collectively called the *advice*, which reside in the central analysis process and which, in the case of StackFul, are used to simultaneously perform concrete and symbolic execution of the application. Specifically, the advice wraps all values produced during the execution of the program into tuples containing both the *concrete* value and the corresponding *symbolic* value's representation. To this end, at certain join points in the program such as method invocations, binary and unary expressions, literal values, variables etc., the instrumented code of the client and server calls the advice remotely.

Listing 4 illustrates some of these advice methods generated by Aran when instrumenting lines 15 to 18 of listing 3. Aran transforms the if-statement **if** (right === 0) into a series of calls to the read, primitive, binary, and test advice methods. These methods are used to inspect and possibly modify the default behaviour when respectively reading the variable right, accessing the literal value 0, evaluating the binary expression right === 0, and evaluating the predicate expression itself. The advice object is represented by the symbol _.

Listing 5 depicts part of the advice that StackFul registers with Aran-Remote. The primitive function (line 1) is called remotely whenever a *literal* join point is reached, i.e., whenever a literal integer, boolean, or string is encountered in the code. The primitive function wraps the value of the literal into a wrapper object containing a concrete and a symbolic field. Its concrete field is the value of the literal, and its symbolic field indicates a symbolic constant. Value types, such as floating-point numbers and object values, that are not supported by StackFul are represented as wrappers containing an empty symbolic field; here marked as SymEmpty. The binary function (line 10) constructs a similar wrapper. Its concrete field is obtained by performing the binary operation on the concrete field of its operands, while its symbolic field is a compound symbolic expression which keeps track of the origin of its operands. The test function updates the path constraint whenever a conditional expression with a valid symbolic representation is encountered. Finally, the apply

■ **Listing 5** Parts of the advice applied by STACKFUL to perform concrete and symbolic execution simultaneously

```
1  advice.primitive = (primitive) => {
2    concrete: primitive,
3    symbolic: switch (typeof primitive) {
4      case "number": new SymInt(primitive);
5      case "boolean": new SymBool(primitive);
6      case "string": new SymString(primitive);
7      default: new SymEmpty();
8    }
9  };
10 advice.binary = (operator, left, right) => ({
11   concrete: binary(operator, left.concrete, right.concrete),
12   symbolic: new SymArithmeticExp(left.symbolic, new SymString(operator), right.symbolic)
13 });
14 var pathConstraint = [];
15 advice.test = (predicate) => {
16   if (! predicate.symbolic.isSymEmpty()) {
17     pathConstraint.push(predicate.symbolic);
18   }
19   return predicate.concrete;
20 };
21 advice.apply = (closure, context, arguments) => {
22   if (closure.concrete === Math.random) {
23     return { concrete: randomValue(), symbolic: new SymInput() };
24   }
25   …
26 };
```

function (line 21) of the advice takes care of function applications. One source of symbolic input parameters are non-deterministic computations such as Math.random: applying Math.random therefore results in a wrapper where the symbolic field is marked as a SymInput (line 23) and the concrete field can either be a 'real' random value, or a value that was precomputed to satisfy some symbolic constraint. In general, the advice also enables STACKFUL to observe when a message or event-handler is registered, and on which object this handler is registered. In case of a user event, this object is usually some html-element. In case of a message handler, this object is a socket. In either case, STACKFUL bookmarks the object so that the central analysis process can afterwards indirectly execute the handler by respectively dispatching the appropriate user event onto the object or by mocking an incoming message.

### 4.2 Central Analysis Process

The *central analysis process* continuously monitors all information that is relevant for determining which program paths are available on the process under test. This includes information about the symbolic conditions the analysis observes and the various message listeners that are registered. A key advantage of using ARAN-REMOTE is that the central analysis process features synchronous communication between all monitored processes, both client and server, on the one hand, and the central process on the other. An unfortunate side-effect is that this central analysis process may incur a rather large memory and performance overhead.

Note that, when applying intra-process testing of the server, incoming messages must be mocked to test the server's message handling code. During its intra-process phase, STACKFUL treats the server as an event-driven application, where the reception of a mocked message is considered an ordinary event, so intra-process testing of the server proceeds in a manner similar to that described in section 2.2. The payload of these messages is treated as consisting entirely of symbolic input parameters.

Therefore, the program path consists of a sequence of incoming, mocked messages as well as the true or false outcomes of conditional branches which may or may not depend on the values of the message's symbolic input parameters. As the analysis process has information available on the symbolic branch conditions as well as which message listeners have been registered, the analysis knows which paths are available and can communicate this information to the test selector. By controlling the sources of non-deterministic execution, such as e. g., Math.random and the sequence and payload of the messages that must be mocked, the analysis process can steer the execution of the server so it follows the path that was prescribed by the test selector. To this end, the analysis process uses the appropriate advice function to intervene in generating the concrete value for a particular symbolic input parameter whenever the source of non-determinism corresponding to this particular parameter is executed by the server.

The analysis process ensures that the sequence of mocked messages is followed by executing the appropriate message listeners in order. To avoid any race conditions, the analysis triggers the reception of new mocked message only when it has determined that the message handler for the previous message has terminated.

Whenever STACKFUL encounters an error during the execution of the application, it remembers the full path constraint that led to this error.

### 4.3 Test Selector

The *test selector* suggests new program paths to explore in the next test run. To this end, the selector maintains a symbolic execution tree of the application under test, similar to the one depicted in figure 2. This execution tree is extended based on the path constraints that are collected by the central analysis process.

STACKFUL supports two exploration strategies: a brute-force strategy and a more refined one. The former finds new program paths to explore by traversing the tree in a breadth-first manner and by negating components of an existing path to create new paths. The latter exploration strategy is based on the one employed by the SYMJS TESTER [19] and keeps track of which variables are read from and written to by individual event handlers. The selector then forms new program paths by constructing event sequences where the constituent event handlers have been selected to maximise the number of conflicting read-writes on their set of shared variables.

When a new path has been selected, it is provided to an SMT solver that, if the path constraint is satisfiable, computes appropriate values for the symbolic input parameters appearing in the path. The test selector then forwards these results to the central analysis process, which will use them in the next test run. STACKFUL employs the Z3 SMT solver [21] for solving integer and boolean formulae, as well as some limited string [34] and regex formulae. The SMT solver is essential in enabling concolic testers to find values for input parameters that lead to certain paths being covered. Z3 has proved itself as a powerful SMT solver [4].

## 4.4 Inter-process Testing

The previous sections described how STACKFUL executes the intra-process phase of the testing process. In this section, we describe the inter-process phase. Initially, this second phase is similar to the first phase in that the central analysis process and the test selector work in tandem to explore the client side of the web application. As the client is generally event-driven, the path constraints collected by the analysis are once again a combination of symbolic branch conditions and triggered event sequences. As in the first phase, the analysis process intervenes in the execution of the client both when a non-deterministic value is generated, and to ensure that the prescribed event sequence is followed. The inter-process phase can be divided in three stages: exploring the client side code until STACKFUL observes a message being sent to the server, determining whether this message could carry a payload that triggers the server error, and starting a new test run in which the previous run is replayed but the payload of the message is changed.

We describe these stages using listing 6 as an example. In this program, a client generates two random integers (lines 4–5) and sends a message msg to the server if the first number is greater than 10. The server performs an additional check. If the x field in the message's payload equals 15 and the y field equals 1, an error is thrown. Suppose that this error was discovered during the intra-process testing phase. The path constraint leading to this error would be $data.y = 1 \land data.x > 15$, with the additional information that a msg message was mocked to exercise the msg handler.

### 4.4.1 Exploring the Client
This stage of the inter-process phase is similar to the exploration of the server during the intra-process phase. The main goal of this stage is to discover a program path that results in a message being sent. If this happens, the analysis process checks whether this message can result in a previously discovered error to be reached on the server side. If any such errors are found, STACKFUL proceeds to the next stage. If a test run finishes without a message being sent, STACKFUL starts a new test run and explores a different path.

In the case of listing 6, one branch of the event handler results in a message being sent. Suppose that at some point a test run generates the random integers 11 and 3 for x and y respectively so that the message at line 7 is sent, with a payload of 12 for data.x and 3 for data.y. STACKFUL remembers the path constraint for the current test run, $x > 10$, and proceeds to the next stage.

■ **Listing 6** Client and server code of a simple application

```
1  // Client side code
2  …
3  document.addEventHandler("click", function (e) {
4    var x = generateRandomInt();
5    var y = generateRandomInt();
6    if (x > 10) {
7      socket.emit("msg", {x: x + 1, y: y });
8    }
9  }
10 // Server side code
11 …
12 socket.on("msg", function (data) {
13   if (data.y === 1 && data.x > 15) {
14     throw new Error();
15   }
16 }
```

#### 4.4.2 Considering the Message's Payload

The goal of this stage is to determine whether the message being sent can carry a concrete payload such that both the current, client side path constraint is satisfied (so that the message will indeed be sent) and the server side path constraint corresponding to that error is also satisfied (so that this payload will lead to the expected error). If this is not possible, the current test run proceeds, but STACKFUL returns to the previous stage until it observes another message being sent. If it is possible, STACKFUL continues with this stage.

Returning to our example, STACKFUL has observed a message msg being sent to the server. A message of this type could result in the error at line 14 being triggered. At this point, STACKFUL must employ the message payload to *synchronise* the client side path constraint with the server side path constraint that led to the error. The server path, $data.y = 1 \land data.x > 15$ employed two symbolic input parameters, $data.x$ and $data.y$ to mock the message payload. The client path defines its own constraints on parts of the payload, namely $x > 10$. Both paths must be joined together by explicitly including constraints that equate all mocking symbolic inputs with their actual symbolic values: $x + 1 = data.x \land y = data.y$. The full, synchronised path constraint is therefore $x > 10 \land x+1 = data.x \land y = data.y \land data.y = 1 \land data.x > 15$. Although the current, concrete values for x and y, 11 and 3 respectively, will not result in the server error being thrown, STACKFUL can feed the full path constraint to an SMT solver to find values for x and y, e.g., 18 and 1, such that the full constraint is satisfied. STACKFUL then proceeds to the final stage.

#### 4.4.3 Replaying the Test Run

STACKFUL starts a new test run in which it uses the values that were computed for the appropriate symbolic parameters in the last stage to verify whether this new run succeeds in reproducing the expected error. If it does, the error is marked as a *high-priority* error, since it has been proven that the client can be exercised in such a way that an error is produced on the server side. STACKFUL afterwards returns to the first stage to resume exploring the client, until its test budget is exhausted. Any remaining unmarked errors are then automatically classified as *low-priority* errors.

In the case of listing 6, it can be observed that the previously computed values for x and y indeed result in the server error being re-triggered. This last stage also gives STACKFUL the opportunity to produce an exact error message detailing how it arrived at this error.

Note that it is generally impossible to *prove* that an error is of only low priority. Given an infinite number of program paths available on the client, STACKFUL can only ascertain that none of the program paths that were explored within a particular budget led to the rediscovery of a previously discovered error.

### 4.5 Conclusion

In order to apply intra-process and inter-process testing for full-stack applications, STACKFUL must overcome the following challenges:

1. It must mock message sends on the server side in order to find errors that are triggered by receiving a particular message from the client. The payload of these messages is mocked by having it take the form of symbolic input values.
2. It must generate appropriate sequences of events on the client side in order for messages to be sent to the server.
3. The inter-process phase must synchronise the client and the server side paths so that constraints on the server side are also applicable to the data generated at the client side.

The first and second challenge are addressed by the exploration strategy. The third challenge is overcome by inserting additional constraints into the concatenation of the client and server side paths. These constraints equate the symbolic input values, used to mock the message payload, with their actual symbolic values.

## 5 Formal Presentation of STACKFUL

To detail the intra- and inter-process testing phases of STACKFUL, we provide a formal model of the approach for a minimalistic language that includes all language features relevant to full-stack JavaScript web applications, such as event handlers and cross-process message sends. Besides conveying how communication between client and server processes affects the internal state maintained by STACKFUL, the model also illustrates the mocking of events and messages, the triggering of event sequences, and the creation of path conditions. We list the most relevant evaluation rules in this section and refer to the appendix for the complete overview.

Figure 8 describes the syntax of the language. Its primitive values consist of integers, booleans, and closures. Atomic expressions are evaluated in a single evaluation step. They consist of constants, lambdas, variables, input-expressions which generate a non-deterministic number when evaluated, binary expressions, and register-expressions for registering a closure as the handler for messages of a given type. The full set of expressions also includes let- and if-expressions, function applications, and message sends via the send-expression. register- and send-expressions correspond respectively to calls to the socket.on, and the socket.emit method in section 3. Registered closures also serve the purpose of acting as event handlers during the inter-process phase. Recall that STACKFUL mocks messages in the intra-process phase and that it mocks user events during the inter-process phase, so the model does not distinguish between the corresponding handler types. For simplicity, we assume that a server in this language does not include any send-expressions.

In the remainder of this section, we define a set of small-step evaluation rules that stipulate how STACKFUL would perform a single test run of an application. These evaluation rules operate on a CESK-machine-like [13] representation of the state of STACKFUL. Figure 9 defines the state space. Overlines in this figure denote sequences of elements. We use the :-operator to concatenate sequences or to conjoin individual elements, and we use $\epsilon$ to represent an empty sequence. A state $\varsigma$ consists of: the current expression being evaluated, the current lexical environment which maps variables to addresses, the store which maps addresses to values, and the continuation stack consisting entirely of continuation frames for let-expressions.

A value in the language consists of a concrete and a symbolic value. A concrete value $v_c$ is either an integer, a boolean, or a closure. A symbolic value $v_s$ can either be a literal integer or boolean, a unique input parameter labelled with an identifier (i.e., a symbol), the **empty** value (in case the concrete value cannot be represented symbolically), or a symbolic representation of a binary expression. Note that the store maps to pairs of a concrete and a symbolic value, rather than just either a concrete or a symbolic value.

The state furthermore features a path constraint, a (possibly empty) sequence of precomputed inputs, and a (possibly empty) sequence of preselected message handlers. As explained in section 2, in an event-driven program, the program path includes both conditional expressions as well as event handlers. The path constraint consists of both the symbolic representations of branch conditions encountered while executing a program and closures that were registered as message handlers. The sequence of inputs represents the precomputed concrete values that will be assigned to the symbolic input parameters encountered during a particular test run. The sequence of message handlers represents the (message or event, depending on the testing phase) handlers selected by STACKFUL to be executed successively. Whenever a particular handler has been completely executed, STACKFUL moves on to the next handler, until all handlers have been executed.

We do not model how STACKFUL moves from one test run to the other, nor how it transitions from the intra-process to the inter-process phase. The former involves selecting the sequence of inputs and of handlers to be triggered in the next run. Both sequences are computed by observing the path constraints that were collected over the previous test runs. The exact mechanism for selecting them depends on the exploration strategy employed by the tester. Instead of modelling this mechanism, we assume the existence of an external driver which provides both sequences and also observes any error detected by STACKFUL. The transition between the intra-process

$$
\begin{aligned}
c \in \textit{Constants} &::= \quad i \mid b \\
lam \in \textit{Lam} &::= \quad \lambda x.\, e \\
f, ae \in \textit{Atom} &::= \quad c \mid lam \mid x \mid \textsf{input} \mid \\
&\qquad lam \mid ae \oplus ae \mid \textsf{register}\ m\ lam \\
x \in \textit{Var} &::= \quad \text{(a set of identifiers)} \\
m \in \textit{HandlerType} &::= \quad \text{(a set of identifiers)} \\
e \in \textit{Exp} &::= \quad \textsf{let}\ x = e\ \textsf{in}\ e \\
&\quad \mid\ f\ ae \\
&\quad \mid\ ae \\
&\quad \mid\ \textsf{if}\ ae\ \textsf{then}\ e\ \textsf{else}\ e \\
&\quad \mid\ \textsf{send}\ m\ ae
\end{aligned}
$$

■ **Figure 8** The syntax of the minimalistic language and the inter-process takes place after a predetermined number of test runs have been completed, and can hence be modelled trivially.

The evaluation rules for STACKFUL are split in evaluation rules for atomic expressions and for the other expressions.

## 5.1 Evaluating Atomic Expressions

Atomic expressions are evaluated via the atomic evaluation function $\mathscr{A}$, defined as

$$\mathscr{A} = \textit{Atom} \times \textit{Env} \times \textit{Store} \times \textit{PC} \times \textit{Inputs} \rightarrow \mathscr{V} \times \textit{PC} \times \textit{Inputs}$$

and listed in figure 10. This function takes as input an atomic expression, an environment and a store, a path constraint and a sequence of precomputed inputs. It returns a value, the possibly updated path constraint, and the possibly updated inputs sequence.

Evaluating a constant produces a value of which the concrete component consists of the constant itself and the symbolic component corresponds to a *lifting* of the concrete value to the symbolic domain according to the lift function ↑. Lifting an integer $i$ produces the symbolic value **int($i$)**, lifting a boolean produces **bool($b$)**. Evaluating a lambda produces a closure, with the symbolic component being **empty** as closures cannot be represented symbolically. A variable is evaluated by retrieving its address from the environment and looking up this address in the store.

Evaluating an input-expression results in a non-deterministic value. The symbolic component of this value is always a new, unique symbolic input parameter $in_{id}$. The concrete component is either the first value $v_c$ in the sequence of precomputed input values, or it is a purely random number if this sequence is empty. In case of the former,

$$
\begin{aligned}
clo \in Closure &::= \mathbf{clo}(lam, \rho) \\
h \in Handler &::= \mathbf{handler}(clo, m) \\
id \in \mathbb{N} &::= \text{(an infinite set of identifiers)} \\
a \in Addr &::= \text{(an infinite set of identifiers)} \\
v \in \mathcal{V}_c &::= i \mid b \mid clo \\
v_s \in \mathcal{V}_s &::= \mathbf{int}(i) \mid \mathbf{bool}(b) \mid in_{id} \mid \mathbf{empty} \mid v_s \oplus v_s \\
v \in \mathcal{V} \subseteq \mathcal{V}_c \times \mathcal{V}_s &::= \langle v_c, v_s \rangle \\
\rho \in Env &= Var \rightarrow Addr \\
\sigma \in Store &= Addr \rightarrow \mathcal{V} \\
\kappa \in Kont &::= \mathbf{letk}(a, e, \rho) \\
\overline{\kappa} \in KStack &= \overline{Kont} \\
ctrt \in Constraint &::= v_s \mid h \\
pc \in PC &= \overline{Constraint} \\
\overline{\iota} \in Inputs &= \overline{\mathcal{V}_c} \\
\overline{h} \in Handlers &= \overline{Handler} \\
\varsigma \in \Sigma &= Exp \times Env \times Store \times KStack \times \\
&\quad PC \times Inputs \times Handlers
\end{aligned}
$$

■ **Figure 9** The state space of STACKFUL

$\mathcal{A}$ returns the remainder $\overline{\iota}$ of the sequence alongside the non-deterministic value. In case of the latter, it returns the empty sequence $\epsilon$.

A register-expression is evaluated by wrapping the closure and the handler type into a **handler** and appending this handler to the path constraint. The boolean true is returned to signal the lambda successfully being registered. The registration is included as part of the path constraint because a handler might only be registered conditionally. Before the start of a new test run, the external driver must therefore consider the path constraint when determining which sequence of handlers to select.

The atomic evaluation rule A-BINARY stipulates that a binary expression is evaluated by successively applying $\mathcal{A}$ to the left and to the right operand. The concrete component of the resulting value is computed by simply applying the operator. In most cases, the symbolic component is simply a literal representation of this operator being applied to the symbolic values of the operands. Depending on the SMT solver that is used [5], however, some expressions involving non-linear binary operators (e.g., modulo) cannot be represented symbolically as they cannot be solved. Concolic testers mitigate this problem via *concretisation*: the concrete result is lifted directly to the symbolic domain.

$$\mathscr{A}(c, \rho, \sigma, pc, \bar{\iota}) = \langle \langle c, \uparrow c \rangle, pc, \bar{\iota} \rangle$$

$$\mathscr{A}(\lambda x.\, e, \rho, \sigma, pc, \bar{\iota}) = \langle \langle \mathbf{clo}(\lambda x.\, e,\, \rho), \mathbf{empty} \rangle, pc, \bar{\iota} \rangle$$

$$\mathscr{A}(x, \rho, \sigma, pc, \bar{\iota}) = \langle \sigma(\rho(x)), pc, \bar{\iota} \rangle$$

$$\mathscr{A}(\mathbf{input}, \rho, \sigma, pc, v_c : \bar{\iota}) = \langle \langle v_c, in_{id} \rangle, pc, \bar{\iota} \rangle \quad \text{With } id \text{ a new, unique identifier}$$

$$\mathscr{A}(\mathbf{input}, \rho, \sigma, pc, \epsilon) = \langle \langle i_r, in_{id} \rangle, pc, \epsilon \rangle \quad \text{With } i_r \text{ a random number and } id \text{ a new, unique identifier}$$

$$\mathscr{A}(\mathbf{register}\; m\; lam, \rho, \sigma, pc, \bar{\iota}) = \langle \langle \mathbf{true}, \mathbf{bool}(\mathbf{true}) \rangle, pc : \mathbf{handler}(\mathbf{clo}(lam, \rho), m), \bar{\iota} \rangle$$

A-Binary
$$\frac{\langle \langle v_{c1}, v_{s1} \rangle, pc', \bar{\iota}' \rangle = \mathscr{A}(ae_1, \rho, \sigma, pc, \bar{\iota}) \qquad \langle \langle v_{c2}, v_{s2} \rangle, pc'', \bar{\iota}'' \rangle = \mathscr{A}((ae_2, \rho, \sigma, pc', \bar{\iota}')}{\mathscr{A}(ae_1 \oplus ae_2), \rho, \sigma, \bar{\iota}) = \langle \langle v_{c1} \oplus v_{c2}, v_s \rangle, pc'', \bar{\iota}'' \rangle}$$

With $v_s$ equal to $v_{s1} \oplus v_{s2}$ if $v_{s1} \oplus v_{s2}$ can be modelled symbolically, or equal to $\uparrow (v_{c1} \oplus v_{c2})$ if it cannot

■ **Figure 10** The atomic evaluation function $\mathscr{A}$

### 5.2 Evaluating Non-atomic Expressions

Figure 11 lists a selection of the evaluation rules for non-atomic expressions. These evaluation rules take a state $\varsigma$ as input and return either a **next** if the test run can proceed with a new state, a **fail** if the test run terminated because an error was encountered, or a **stop** if the test run was stopped either because StackFul executed all preselected handlers or because it found a message send that will result in a high-priority error being triggered. In case of a **fail**, evaluation stops and returns the current path constraint as well as the handler type of the handler that was being executed.

Rule E-Let stipulates that let-expressions are evaluated by allocating a new, unique address and creating a new environment $\rho'$ where the variable is bound to this address. StackFul pushes a new **letk** continuation frame, to evaluate the let-body-expression once the let-value-expression has been evaluated. Note that let-expressions are the only type of expressions that result in a frame being pushed onto the continuation stack.

E-PopContinuation concerns atomic expressions. It uses the evaluation function $\mathscr{A}$ to evaluate this expression to a value $v$, and it considers the continuation stack to determine where this value must flow to. Since the stack consists entirely of **letk** frames, StackFul proceeds by assigning $a$, saved in the frame, to $v$ in the store and continuing with the let-body-expression.

E-IfTrue demonstrates how the path constraint is updated when a branch condition is encountered. It specifies that if the predicate of an if-expression evaluates to true, the path constraint is updated by appending the symbolic component of the predicate value to the path constraint. The inverse of this rule, E-IfFalse, is listed in the appendix.

E-HandlerWithInput describes that a preselected handler is triggered when StackFul has reached an atomic expression with an empty continuation stack. As this

E-Let

$$\frac{a = \text{ALLOC}() \quad \rho' = \rho[x \mapsto a]}{\langle \text{let } x = e_1 \text{ in } e_2, \rho, \sigma, \kappa, pc, \overline{\iota}, \overline{h} \rangle \to \textbf{next}(\langle e_1, \rho, \sigma, \textbf{letk}(a, e_2, \rho') : \kappa, pc, \overline{\iota}, \overline{h} \rangle)}$$

E-PopContinuation

$$\frac{\mathscr{A}(ae, \rho, \sigma, pc, \overline{\iota}) = \langle v, pc', \overline{\iota}' \rangle}{\langle ae, \rho, \sigma, \textbf{letk}(a, e', \rho') : \kappa, pc, \overline{\iota}, \overline{h} \rangle \to \textbf{next}(\langle e', \rho', \sigma[a \mapsto v], \kappa, pc', \overline{\iota}', \overline{h} \rangle)}$$

E-IfTrue

$$\frac{\mathscr{A}(ae, \rho, \sigma, pc, \overline{\iota}) = \langle \langle \text{true}, v_s \rangle, pc', \overline{\iota}' \rangle}{\langle \text{if } ae \text{ then } e_1 \text{ else } e_2, \rho, \sigma, \kappa, pc, \overline{\iota}, \overline{h} \rangle \to \textbf{next}(\langle e_1, \rho, \sigma, \kappa, pc' : v_s, \overline{\iota}', \overline{h} \rangle)}$$

E-HandlerWithInput

$$\frac{a = \text{ALLOC}() \quad \rho'' = \rho'[x \mapsto a] \quad \sigma' = \sigma[a \mapsto \langle v_c, in_{id} \rangle]}{\langle ae, \rho, \sigma, \epsilon, pc, v_c : \overline{\iota}, \textbf{handler}(\textbf{clo}(\lambda x.\, e', \rho'), m') : \overline{h} \rangle \to \textbf{next}(\langle e', \rho'', \sigma, \epsilon, pc, \overline{\iota}, \overline{h} \rangle)}$$

With $id$ a new, unique identifier.

E-Error

$$\frac{\mathscr{A}(ae, \rho, \sigma, pc, \overline{\iota}) = \text{\textreferencemark} \quad m = \text{GetCurrentHandlerType}()}{\langle ae, \rho, \sigma, \kappa, pc, \overline{\iota}, \overline{h} \rangle \to \textbf{fail}(pc, m)}$$

E-SendSatisfiable

$$\frac{\mathscr{A}(ae, \rho, \sigma, pc, \overline{\iota}) = \langle \langle v_c, v_s \rangle, pc', \overline{\iota}' \rangle \quad \exists pc'' \in \theta_{server}(m) : pc' : (in_{id} = v_s) : pc'' \text{ is satisfiable}}{\langle \text{send } m\ ae, \rho, \sigma, \kappa, pc, \overline{\iota}, \overline{h} \rangle \to \textbf{stop}}$$

Where $in_{id}$ is the symbolic input parameter assigned to the mocked message handler that caused the error.

**Figure 11** Evaluating non-atomic expressions

would ordinarily signal the end of the program's execution, STACKFUL pops the first handler from the sequence and proceeds by evaluating the body of the closure. Since the parameter of the closure is a non-deterministic input parameter, it is assigned the first value $v_c$ from the sequence of precomputed inputs, if this sequence is non-empty, similar to how the atomic evaluation function evaluated input-expressions. We refer to the appendix for cases where this sequence of inputs is empty (E-HANDLERNOINPUT) or where no more preselected handlers are available (E-NOMOREHANDLERS).

E-ERROR describes how STACKFUL handles the case where, during the execution of a mocked handler, atomic evaluation results in an error, represented here as the symbol $\lightning$, e.g., because an undefined variable was read or a binary operator was applied to incompatible types. We employ the auxiliary function GETCURRENTHANDLERTYPE, not modelled here, to find the type of the preselected handler currently being executed. STACKFUL wraps the current path constraint as well as the handler type $m$ of the handler currently being mocked in a **fail**. For brevity, we assume that the external driver holds a collection of errors $\theta$ defined as a map from a handler type to a collection of path constraints.

E-SENDSATISFIABLE models the second stage of inter-process testing (section 4.4.2) and describes the evaluation of a send-expression on the client side. We assume that the external driver provides the error collection $\theta_{server}$ that was gathered after intra-process testing of the server was completed.

The goal is to find a server error, previously reported via the E-ERROR rule during intra-process testing, which would become reachable because of the message send. The error can only be reachable if both the current path constraint $pc'$ is true (so that this message is sent in the first place) and the server side path constraint $pc''$ is also true (so that upon arriving, the message leads to the error). Furthermore, messages carry payloads, represented as the atomic expression $ae$ in this rule. While this payload was mocked with a symbolic input parameter during the intra-process testing phase, it now receives an actual value by applying the atomic evaluation function on the payload. This value must explicitly be bound in the path constraint to the original input parameter[2] of the mocked message handler whose execution led to the error during the intra-process phase. Any check performed by the server on the mocked input would have appeared as a constraint in $pc''$, while any check that appeared on the message payload would have appeared in $pc'$. By explicitly equating the mocked input with the actual value, any check applied by one side also becomes applicable to the other. STACKFUL therefore joins $pc'$ and $pc''$ together via the equality $in_{id} = v_s$. If the concatenation of both path constraints is satisfiable, the server error can be reproduced via this message send, so the test run stops.

---

[2] We do not model how the original input parameter $in_{id}$ is found, but it appears in the path constraint $pc''$.

**Table 1** Characteristics of the web applications considered in our study

| Benchmark | LOC | # of Branches | WMC | # of Message and Event Handlers |
|---|---|---|---|---|
| Calculator | 126 | 16 | 18 | 16 |
| Chat | 288 | 39 | 27 | 9 |
| Game of Life | 214 | 64 | 32 | 10 |
| Simple Chat | 45 | 4 | 6 | 2 |
| Slack Mockup | 662 | 75 | 20 | 13 |
| TOHacks | 144 | 22 | 20 | 5 |
| Totems | 145 | 9 | 14 | 5 |
| Whiteboard | 126 | 16 | 14 | 3 |

# 6 Evaluation

The goal of our evaluation is to compare the inter-process and intra-process (baseline) approaches to concolic testing of full-stack JavaScript web applications. Specifically, we measure how capable inter-process testing is at discerning low-priority from high-priority errors. We define the following research questions:

- **RQ1:** *How many high-priority server errors are correctly classified by StackFul as being of high priority?*
- **RQ2:** *Are there any instances of StackFul incorrectly classifying low-priority server errors as being of high priority?*
- **RQ3:** *How many test runs of inter-process testing does StackFul require to reproduce a high-priority error?*

The context of the study consists of eight different programs (Calculator, Chat, Game of Life, Slack Mockup, Simple-chat, TOHacks, Totems, and Whiteboard). The Calculator program is the motivating example introduced in section 3. Whiteboard and Chat are demo applications[3] for the Socket.IO library. Game of Life, TOHacks, Totems, and Simple-chat were retrieved from a software gallery[4] featuring applications built with the Socket.IO library. Slack Mockup is a project that mimics some of functionalities of the Slack communication platform. Apart from the Calculator application, all programs maintain both a client and a server side state, which is manipulated by the event and message handlers. Table 1 reports the main characteristics of each web application: the number of lines of code, the number of branches in the code, the Weighted Methods per Class (WMC) metric, and the number of message and event handlers that are registered in the application.

To evaluate the inter-process approach, we introduced additional synthetic pairs of subsuming checks in both the client and server side for the Game of Life, Simple Chat, Totems, and Whiteboard programs. These subsuming checks, similar to the

---

[3] https://github.com/socketio/socket.io/tree/master/examples, last accessed 2020-10-01.
[4] https://devpost.com/software/built-with-socket-io, last accessed 2020-10-01.

■ **Listing 7** Part of the server code for one of the two Calculator variants, with an injected synthetic fault.

```
case "+": {
  "ERROR: INJECTED SERVER ERROR";
  {
    result = left + right;
    break;
  }
case "-": result = left - right; break;
```

client and server side checks featured in the Calculator example, conform to the following pattern: the client checks whether a condition on a value holds, and, if it does, sends a message pertaining to this value to the server. The server then checks a similar, i.e., subsuming, condition as the one performed by the client. Examples of such checks include whether or not the coordinates of a mouse click fall within or outside a certain frame, or whether certain words in a chat application should be censored instead of being broadcasted. These checks effectively create regions of dead code on the server,[5] as, given this particular client-server configuration, these conditions should never be false.

Having introduced these checks, we automatically and randomly inject synthetic "ERROR: INJECTED SERVER ERROR" faults in the server side of the resulting programs. This because the collected benchmark programs contain few faults by themselves. Every if-branch in the program has the same probability of being injected by the fault injection process. Faults injected in the server that fall within the aforementioned dead code region are considered *low-priority* errors, while faults injected outside of these regions are *high-priority*. Listing 7, which corresponds to lines 12–13 of listing 3, depicts an "ERROR: INJECTED SERVER ERROR" fault being injected at line 2.

To increase the generalisability of our study, we applied the fault injection process to create two variants of the server side code of each benchmark application, except for the Simple-chat application, as the small size of this application's code base rendered the differences between both variants negligible. To answer the research questions, we ran StackFul on all 15 resulting applications. For the first phase, StackFul was allocated a test budget of 250 test runs to ensure all server side errors (both high- and low-priority errors) were found. In the second phase, a test budget of 500 test runs was allocated to rediscover these errors.

Table 2 reports on the classification of high- and low-priority errors for each application. For both categories, the table specifies the total number of errors of that category that were injected into the server side of that application (*# of Faults*), and the number of these errors that StackFul correctly labelled as belonging to that category (*Correctly Classified*). We manually verified for each application whether the classification generated by StackFul was correct.

---

[5] Note that, since these checks depend on the value of the payload, static dead code checks do not suffice for discovering these regions.

■ **Table 2** Classification in high- and low-priority errors.

| Application | High-Priority | | Low-Priority | |
| --- | --- | --- | --- | --- |
| | # of Faults | Correctly Classified | # of Faults | Correctly Classified |
| CALCULATOR I | 5 | 3 | 2 | 2 |
| CALCULATOR II | 6 | 6 | 1 | 1 |
| CHAT I | 0 | 0 | 1 | 1 |
| CHAT II | 1 | 0 | 2 | 2 |
| GAME OF LIFE I | 1 | 1 | 1 | 1 |
| GAME OF LIFE II | 2 | 1 | 2 | 1 |
| SIMPLE CHAT | 0 | 0 | 1 | 1 |
| SLACK MOCKUP I | 3 | 0 | 3 | 2 |
| SLACK MOCKUP II | 4 | 1 | 1 | 1 |
| TOHACKS I | 3 | 3 | 0 | 0 |
| TOHACKS II | 2 | 2 | 0 | 0 |
| TOTEMS I | 2 | 2 | 0 | 0 |
| TOTEMS II | 0 | 0 | 2 | 2 |
| WHITEBOARD I | 2 | 2 | 2 | 2 |
| WHITEBOARD II | 5 | 5 | 3 | 3 |

We answer the three research questions over the following sections. In section 7, we go into more details on the answers that are offered here.

### 6.1 How many high-priority server errors are correctly classified by STACKFUL as being of high priority?

All of the applications combined contain a total 36 high-priority server errors. STACK-FUL is able to correctly classify 26 of these. CALCULATOR I, CHAT II, GAME OF LIFE II and both versions of SLACK MOCKUP contain misclassified high-priority errors. For the first three applications, these errors were uncovered during the intra-process phase but were not reproduced during the inter-process phase, even though they were reachable from the client. In effect, STACKFUL incorrectly labelled these high-priority errors as being of low priority. For the SLACK MOCKUP variants, these errors were also not uncovered during the intra-process phase.

In all cases, these errors are located along program paths that can only be reached from the client when several client side events are triggered in a particular order, and when additional client side conditions are met. This phenomenon was particularly outspoken in SLACK MOCKUP, as some of the event handlers involved are not registered upfront, but only dynamically under certain conditions. STACKFUL therefore first needs to discover a program path where the appropriate handlers are registered before this path can be explored further by triggering the corresponding sequence of events. In general, it is conceivable that STACKFUL would have correctly classified the errors for all of these applications given a larger test budget. Practical concerns, such as the long (up to five hours for the budget of $250 + 500 = 750$ combined test runs) running time of each benchmark and the memory footprint of ARAN-REMOTE's approach to dynamic analysis of distributed applications, necessitate a limit on the test budget.

■ **Table 3** The number of inter-process test runs STACKFUL required to classify each of the high-priority errors it found

| Error | # of Test Runs | Error | # of Test Runs |
| --- | --- | --- | --- |
| CALCULATOR I: Error 1 | 33 | SLACK MOCKUP II: Error 1 | 240 |
| CALCULATOR I: Error 2 | 33 | SLACK MOCKUP II: Error 1 | > 500 |
| CALCULATOR I: Error 3 | 449 | SLACK MOCKUP II: Error 1 | > 500 |
| CALCULATOR I: Error 4 | > 500 | SLACK MOCKUP II: Error 1 | > 500 |
| CALCULATOR I: Error 5 | > 500 | TOHACKS I: Error 1 | 1 |
| CALCULATOR II: Error 1 | 33 | TOHACKS I: Error 2 | 1 |
| CALCULATOR II: Error 2 | 33 | TOHACKS I: Error 3 | 12 |
| CALCULATOR II: Error 3 | 33 | TOHACKS II: Error 1 | 1 |
| CALCULATOR II: Error 4 | 33 | TOHACKS II: Error 2 | 1 |
| CALCULATOR II: Error 5 | 33 | TOTEMS I: Error 1 | 1 |
| CALCULATOR II: Error 6 | 341 | TOTEMS I: Error 2 | 1 |
| CHAT II: Error 1 | > 500 | WHITEBOARD I: Error 1 | 4 |
| GAME OF LIFE I: Error 1 | 7 | WHITEBOARD I: Error 2 | 9 |
| GAME OF LIFE II: Error 1 | 8 | WHITEBOARD II: Error 1 | 3 |
| GAME OF LIFE II: Error 2 | > 500 | WHITEBOARD II: Error 2 | 3 |
| SLACK MOCKUP I: Error 1 | > 500 | WHITEBOARD II: Error 3 | 9 |
| SLACK MOCKUP I: Error 2 | > 500 | WHITEBOARD II: Error 4 | 15 |
| SLACK MOCKUP I: Error 3 | > 500 | WHITEBOARD II: Error 5 | 15 |

We conclude that, for the given applications, STACKFUL is able to classify the majority of high-priority errors correctly without exhausting its test budget.

### 6.2 Are there any instances of STACKFUL incorrectly classifying low-priority server errors as being of high priority?

As can be observed from table 2, the applications total a number of 21 low-priority server errors. STACKFUL misclassifies only two of these, one for the GAME OF LIFE II application and one for the SLACK MOCKUP I application. In both cases, these errors were only misclassified because they were not discovered during the intra-process testing phase in the first place. However, STACKFUL did not reproduce any of the low-priority errors during its inter-process testing phase. There were hence no instances of low-priority errors being misclassified as being of high priority.

### 6.3 How many test runs of inter-process testing does STACKFUL require to reproduce a high-priority error?

Table 3 specifies for each high-priority error that was injected into an application how many test runs STACKFUL required to reproduce this error, if at all, during its inter-process testing phase. Errors that were misclassified, i.e., the errors in CALCULATOR I and CHAT II, must have required *at least* 500 test runs before they are reached. Note that we assume that STACKFUL would eventually reproduce a high-priority error in its inter-process testing phase, given an infinite test budget.

It can be observed that these errors fall into two categories: those errors which StackFul can reproduce in a rather small number of test runs (i.e., 33 runs or less), and those for which it requires many more test runs. In section 7, we hypothesise a likely explanation for this stark contrast.

### 6.4 Threats to Validity

We identify the threats to construct and external validity of our empirical evaluation.

#### 6.4.1 Construct Validity

We have developed StackFul to investigate the strengths and weaknesses of inter-process testing. It is worth noting that StackFul lacks some optimisations that are generally included in automated testers such as state-of-the-art concolic testers. These optimisations would certainly have impacted the effectiveness of both the intra- and the inter-process testing phase with regards to the number of test runs required to respectively discover *any* server error, and to classify an error as a high-priority error. In the particular set of applications that we investigated, StackFul discovered most injected server errors during the intra-process phase but was unable to reproduce some of these errors during the inter-process phase. We assert that additional optimisations to StackFul would have prevented high-priority errors from being misclassified. We come back to this statement in section 7. We have attempted to mitigate this thread by setting the inter-process test budget at 500 test runs, and by including the exploration strategy employed by the SymJS tester [19] alongside the brute-force exploration strategy.

#### 6.4.2 External Validity

The main threat to the external validity of our evaluation is the limited number of programs on which we conducted it. We partially mitigated this threat by creating variants of the original programs in which we automatically and systematically inject faults to avoid possible bias. Replicating the study on larger open source web applications is part of our future agenda, and will among others require an automated means for discovering full-stack JavaScript applications on GitHub.

## 7 Discussion

The results of section 6.1 and 6.2 indicate that StackFul is generally able to correctly categorise discovered errors. However, the results also highlight two related weaknesses in StackFul's approach. First, if StackFul requires more inter-process test runs than its test budget allows to reproduce a server error, the error is automatically classified as a low-priority error, even if the error is reachable in the given client-server configuration. All of the misclassified high-priority errors that were described in section 6.1 and 6.3 suffer from this effect. Second, StackFul can only assume that an error is of low priority in a particular client-server configuration if the error cannot be reproduced over the span of the inter-process testing phase. Therefore, StackFul must always exhaust its entire test budget to classify a collection of low-priority errors. We discuss these weaknesses separately.

### 7.1 Misclassification of High-priority Errors

The first weakness can be mitigated by employing more sophisticated search strategies during the inter-process testing phase. Investigating the number of required test runs as described in section 6.3 more closely, it becomes apparent that the inter-process phase performs at its worst when attempting to reproduce server errors that are only reachable from the client by following a specific sequence of client side events, while satisfying additional client side branch conditions. This effect is particularly pronounced in the case of SLACK MOCKUP, which is not only the largest benchmark but which also features the most intricate configuration of event handlers. STACKFUL features two strategies for exploring event-driven applications: a simple brute-force strategy and another strategy that aims to maximise read-write conflicts between event handlers. However, neither strategy is able to overcome this issue.

More efficient exploration strategies might help to reduce the time STACKFUL needs to reproduce an error, and therefore prevent STACKFUL from misclassifying high-priority errors. Strategies that are better suited for testing highly event-driven applications might yield improved results. Several such exploration strategies for event-driven and message-driven applications have been described, such as dCute [28], which generates event sequences based on a partial ordering of the events, or CONTEST [1], which identifies subsuming event sequences. Another technique for prioritising event sequences or program paths that lead the client to communicate with the server is to initialise the test selector with program-specific *path prefixes* [3, 24]. These are programmer-defined sequences of events that are guaranteed to lead the tester to exercise a desired part of the application. All event sequences that are generated by the test selector must start with one of these prefixes. We believe that implementing these techniques will be essential in enabling STACKFUL to scale to larger applications.

### 7.2 Efficiency of Classifying Low-priority Errors

The efficiency of STACKFUL's inter-process testing phase would be improved if STACKFUL could *prove* that a low-priority error is indeed unreachable in a given client-server configuration. This would prevent STACKFUL from having to exhaust its entire inter-process test budget in order to classify low-priority errors. However, to prove that a server error is unreachable from a given client, it must be demonstrated that no program path in the entire web application starts from an entry-point in the client (such as an event handler) and reaches the given error. In practice, even just the client side of the application is likely to have an infinite number of program paths, so exploring all of these to determine non-reachability is not possible.

One option would be to compose function summaries [8, 14, 17] of the client side event handlers. Once all summaries have been constructed, they could be used to determine whether any sequence of events can result in a particular server error being reached. If not, the server error is a low-priority error. Note that this would also solve the first weakness of STACKFUL.

A related idea is the use of *backwards symbolic execution* (BSE) [12, 20], in which a symbolic execution engine executes a program 'backwards' from a particular program statement until it reaches an entry point of the program. However, it is unclear how this technique would fare in the domain of full-stack web applications. A BSE engine would first have to determine whether a particular server error is reachable from a *server* entry point (i.e., a message handler), and then check whether this message handler is reachable from a *client* entry point.

### 7.3 Use Cases for Inter-process Testing

The empirical results of section 6 show that STACKFUL avoided misclassifying low-priority errors. We believe this merits a closer look at potential use cases for this technique. For example, consider a company where the application's code base is tested by STACKFUL overnight. In the morning, developers can be presented with a bug report that suggests which errors are to be addressed first. Providing this additional classification might increase developers' confidence in the testing tool, as evidenced for static bug detectors [18, 23, 25].

Moreover, as the inter-process tester closely mimics how a user would trigger these bugs in practice, the reports produced by this tool could make it easier for developers to diagnose them, as the tester can produce a practical step-by-step guide for their reproduction. We draw an analogy between intra- and inter-process testing on the one hand, and unit- and end-to-end testing on the other. If a certain part of the application is well-covered by unit tests, finding bugs in this part might be done efficiently by running the unit tests. However, end-to-end testing is better suited for mimicking a user's behaviour, or for finding bugs that are the result of a mismatch between individual components.

## 8 Related Work

There has been earlier work on automated white-box testing of JavaScript, and of websites and web applications, both for client side scripts and for server side programs. We also discuss related work on testing distributed programs, as full-stack applications can be considered a form of distributed JavaScript applications. We consider both regular concolic testing tools, more generic symbolic execution frameworks, and other automated testing tools that do not employ any symbolic execution.

SymJS [19] is one of the most complete symbolic execution tools for testing JavaScript programs. SymJS can be used to test both plain JavaScript programs and client side web scripts. It features a string-numeric solver and is able to accurately represent the Document Object Model of the webpage. SymJS also applies various optimisations to reduce the number of states that are observed by the tester, such as state merging and the employment of ITE-expressions. SymJS's event-exploration strategy is one of two strategies that can be used by STACKFUL to test a web application. Kudzu [27] is another symbolic execution framework for JavaScript. It mainly focuses on discovering security vulnerabilities in client side JavaScript code via extensive symbolic string reasoning. Jalangi [30] also performs concolic testing of JavaScript code. Like STACKFUL, it employs program instrumentation to perform the symbolic execution alongside the concrete execution. However, Jalangi only works on plain JavaScript programs and cannot be used to test client side web scripts. Cosette [26] is another recently released symbolic execution framework for JavaScript, where the symbolic execution has been proven to be sound. However, it does not appear to be possible to use Cosette to holistically test a full-stack JavaScript application. Furthermore, Cosette employs static symbolic execution, whereas STACKFUL performs concolic (i.e., dynamic) execution.

ArtForm [33] is a tool for finding bugs in dynamic, form-based websites, where user interactions such as filling in text input fields or clicking buttons may trigger a complex cascade of event handlers. ArtForm has a manual mode, where the user must interact with the website manually, and a concolic testing mode, in which the tool automatically generates inputs to explore new parts of the code. However, ArtForm only focuses on exploring the client side of the website, and does not test the server side. It also does not track the communication between both sides. Craxweb [32] employs the $S^2E$ [7] symbolic execution framework to find security vulnerabilities in server programs automatically. To this end, Craxweb automatically generates HTTP requests containing symbolic data and observes how this data influences the server's execution: e.g., if the HTTP request causes the server to perform an SQL query, an SQL injection may be possible by providing the appropriate concrete values for the symbolic data in the HTTP request. Note that, although the HTTP requests sent by the client can be simulated by Craxweb, the actual client code is not tested.

All of these automated testing tools share the property that they test only one side of the web application. STACKFUL's novelty lies in the fact that it is the first automated white-box tester to holistically test both sides of a full-stack application.

Sen and Agha described an algorithm for applying concolic testing to distributed programs that communicate via messages [28]. Their algorithm is implemented in jCute [29], a concolic tester for Java. The concolic tester not only provides concrete values for the symbolic input parameters of the individual processes, but it also prescribes a specific schedule for the messages that are sent and received by the processes. Prescribing these message schedules is similar to how STACKFUL prescribes event sequences to the various processes of the web application. Like STACKFUL, their algorithm also tracks the symbolic values of the contents of the messages being sent across process boundaries. STACKFUL could therefore be framed as a continuation of Sen and Agha's work, specialised to full-stack web applications. However, these full-stack applications involve their own set of challenges, e.g., automatically simulating user interactions and instrumenting browser scripts.

Client-server communication can also be verified by using static techniques, e.g., by employing session types [11]. As with all static techniques, static types must over-approximate the behaviour of the application. They may therefore incorrectly conclude that certain parts of the server's code are reachable, and are therefore likely to categorise low-priority errors as being of high priority.

## 9 Conclusion

We have presented the first combined intra- and inter-process approach to concolic testing of full-stack JavaScript web applications. In this approach, a tester first finds errors on a server by applying traditional, intra-process testing of the server. In a second phase, the tester explores a particular client-server configuration in an inter-process manner to reproduce these errors. Any error for which it succeeds in doing so is classified as a high-priority error, as end users that access the server via this particular client can trigger the server error. Server errors that the tester cannot reproduce in this configuration are classified as low-priority.

We have developed STACKFUL, which employs this approach, and have evaluated it on a selection of two fault-seeded variants of eight benchmark programs. We compared this approach in terms of how many errors it correctly classifies, and how many test runs it requires to classify high-priority errors.

The results show that the inter-process approach correctly classifies the majority of errors. We deem this classification a major advantage of the inter-process approach. As future work, we plan to optimise STACKFUL's path exploration strategy to reduce the number of test runs required to classify high-priority errors correctly.

**Acknowledgements**   This project was partially supported by the Excellence of Science Project SECO-Assist (O015718F, FWO - Vlaanderen and F.R.S.-FNRS).

## A  Complete Formalisation of STACKFUL

Section 5 presented selected parts of a formalisation of STACKFUL with the aim to illustrate the intra- and inter-process testing phases applied by STACKFUL in more detail. We provide the complete set of small-step evaluation rules here. To this end, we design a minimalistic language that mimics language features essential to full-stack JavaScript applications, such as message and event handlers and message sends. We also allow for non-deterministic computations via input-expressions. Since the syntax of the language (figure 8), the state space of the CESK-machine representing STACKFUL (figure 9), and the evaluation rules for atomic expressions (figure 10) have already been presented in their entirety in section 5, we do not list them again.

Figure 12 lists all evaluation rules for the evaluation of non-atomic expressions. Rules E-APPLICATION, E-IFFALSE, E-HANDLERNOINPUT, and E-NOMOREHANDLES are new; the other rules were previously discussed in section 5.

E-APPLICATION describes how function applications of the form $f\ ae$ are evaluated. As both the function and the argument expression are atomic expressions, they are evaluated with the atomic evaluation function $\mathscr{A}$, with the function expression required to evaluate to a closure. The closure's environment is extended with a binding from the parameter to a new, unique address $a$, and the store is extended with a binding from $a$ to the argument value. Evaluation then proceeds with the function's body.

E-IFFALSE is similar to E-IFTRUE, previously described in section 5. This rule describes the case where the predicate expression evaluates to false. Like E-IFTRUE, the path constraint is extended, but with the *negation* of the symbolic component of the predicate value.

E-HandlerNoInput is similar to E-HandlerWithInput but describes the case where a new handler is mocked and *no* precomputed input is available. Similar to how input-expressions are evaluated when no precomputed inputs are available, a value-pair is generated where the concrete component is $i_r$ a random value, and the symbolic component is a new, unique symbolic input parameter.

E-NoMoreHandlers stipulates that a test run terminates when StackFul reaches an atomic expression with an empty continuation stack and an empty sequence of precomputed handlers. Assuming the evaluation of this atomic expression does not result in an error (cf. E-Error), the test run terminates with a **stop**.

We also define the *inject*-function for creating a new state, given the program expression $e$, the precomputed inputs $\bar{\iota}$, and the preselected handlers $\bar{h}$ as

$$inject(e, \bar{\iota}, \bar{h}) = \langle e, \emptyset, \emptyset, \epsilon, \bar{\iota}, \bar{h} \rangle$$

We do not model how the precomputed inputs and handlers would be selected, but we assume these to be provided by the external driver.

E-LET

$$\frac{a = \text{ALLOC}() \quad \rho' = \rho[x \mapsto a]}{\langle \text{let } x = e_1 \text{ in } e_2, \rho, \sigma, \kappa, pc, \bar{\iota}, \bar{h} \rangle \to \textbf{next}(\langle e_1, \rho, \sigma, \textbf{letk}(a, e_2, \rho') : \kappa \rangle, pc, \bar{\iota}, \bar{h})}$$

E-POPCONTINUATION

$$\frac{\mathscr{A}(ae, \rho, \sigma, pc, \bar{\iota}) = \langle v, pc', \bar{\iota}' \rangle}{\langle ae, \rho, \sigma, \textbf{letk}(a, e', \rho') : \kappa, pc, \bar{\iota}, \bar{h} \rangle \to \textbf{next}(\langle e', \rho', \sigma[a \mapsto v], \kappa, pc', \bar{\iota}', \bar{h} \rangle)}$$

E-APPLICATION

$$\frac{\mathscr{A}(f, \rho, \sigma, pc, \bar{\iota}) = \langle \langle \textbf{clo}(\lambda x.\, e', \rho'), \text{empty} \rangle, pc', \bar{\iota}' \rangle \quad \mathscr{A}(ae, \rho, \sigma, pc', \bar{\iota}') = \langle v, pc'', \bar{\iota}'' \rangle \quad a = \text{ALLOC}()}{\langle f\ ae, \rho, \sigma, \kappa, pc, \bar{\iota}, \bar{h} \rangle \to \textbf{next}(\langle e', \rho'[x \mapsto a], \sigma[a \mapsto v], \kappa, pc'', \bar{\iota}'', \bar{h} \rangle)}$$

E-IFTRUE

$$\frac{\mathscr{A}(ae, \rho, \sigma, pc, \bar{\iota}) = \langle \langle \text{true}, v_s \rangle, pc', \bar{\iota}' \rangle}{\langle \text{if } ae \text{ then } e_1 \text{ else } e_2, \rho, \sigma, \kappa, pc, \bar{\iota}, \bar{h} \rangle \to \textbf{next}(\langle e_1, \rho, \sigma, \kappa, pc' : v_s, \bar{\iota}', \bar{h} \rangle)}$$

E-IFFALSE

$$\frac{\mathscr{A}(ae, \rho, \sigma, pc, \bar{\iota}) = \langle \langle \text{false}, v_s \rangle, pc', \bar{\iota}' \rangle}{\langle \text{if } ae \text{ then } e_1 \text{ else } e_2, \rho, \sigma, \kappa, pc, \bar{\iota}, \bar{h} \rangle \to \textbf{next}(\langle e_2, \rho, \sigma, \kappa, pc' : \neg v_s, \bar{\iota}', \bar{h} \rangle)}$$

E-HANDLERWITHINPUT

$$\frac{a = \text{ALLOC}() \quad \rho'' = \rho'[x \mapsto a] \quad \sigma' = \sigma[a \mapsto \langle v_c, in_{id} \rangle]}{\langle ae, \rho, \sigma, \epsilon, pc, v_c : \bar{\iota}, \textbf{handler}(\textbf{clo}(\lambda x.\, e', \rho'), m') : \bar{h} \rangle \to \textbf{next}(\langle e', \rho'', \sigma, \epsilon, pc, \bar{\iota}, \bar{h} \rangle)}$$

With *id* a new, unique identifier.

E-HANDLERNOINPUT

$$\frac{a = \text{ALLOC}() \quad \rho'' = \rho'[x \mapsto a] \quad \sigma' = \sigma[a \mapsto \langle i_r, in_{id} \rangle]}{\langle ae, \rho, \sigma, \epsilon, pc, \epsilon, \textbf{handler}(\textbf{clo}(\lambda x.\, e', \rho'), m') : \bar{h} \rangle \to \textbf{next}(\langle e', \rho'', \sigma, \epsilon, pc, \bar{\iota}, \bar{h} \rangle)}$$

With *id* a new, unique identifier and $i_r$ a random number

E-ERROR

$$\frac{\mathscr{A}(ae, \rho, \sigma, pc, \bar{\iota}) = \notz \quad m = \text{GETCURRENTHANDLERTYPE}()}{\langle ae, \rho, \sigma, \kappa, pc, \bar{\iota}, \bar{h} \rangle \to \textbf{fail}(pc, m)}$$

E-NOMOREHANDLERS

$$\frac{\mathscr{A}(ae, \rho, \sigma, pc, \bar{\iota}) = \langle v, pc', \bar{\iota}' \rangle}{\langle ae, \rho, \sigma, \epsilon, pc, \bar{\iota}, \epsilon \rangle \to \textbf{stop}}$$

E-SENDSATISFIABLE

$$\frac{\mathscr{A}(ae, \rho, \sigma, pc, \bar{\iota}) = \langle \langle v_c, v_s \rangle, pc', \bar{\iota}' \rangle \quad \exists pc'' \in \theta_{server}(m) : pc' : (in_{id} = v_s) : pc'' \text{ is satisfiable}}{\langle \text{send } m\ ae, \rho, \sigma, \kappa, pc, \bar{\iota}, \bar{h} \rangle \to \textbf{stop}}$$

Where $in_{id}$ is the symbolic input parameter assigned to the mocked message handler that caused the error.

**Figure 12** Evaluating non-atomic expressions

Prioritising Server Side Reachability via Inter-process Concolic Testing

## About the authors

**Maarten Vandercammen** is a Ph.D candidate at the Software Languages Lab of the Vrije Universiteit Brussel. His research is centered around dynamic analysis and automated testing of web applications. Contact him at Maarten.Vandercammen@vub.be.

**Laurent Christophe** is a Ph.D candidate at the Software Languages Lab of the Vrije Universiteit Brussel. His main research interests are the instrumentation and dynamic analysis of JavaScript applications and distributed processes. Contact him at Laurent.Christophe@vub.be.

**Dario Di Nucci** is an assistant professor at the Jheronimus Academy of Data Science. His research is on empirical software engineering, in particular software maintenance and evolution and software testing. To this aim, he applies several techniques such as machine learning, search-based algorithms, and mining of software repositories. He regularly serves as a program committee member of various international conferences (e.g., ESEC/FSE, ICSME, SANER, ICPC), and as referee for various international journals in the field of software engineering (e.g., TSE, TOSEM, EMSE, JSS) and artificial intelligence (e.g., TKDE, Neurocomputing). Contact him at D.DiNucci@tilburguniversity.edu.

**Wolfgang De Meuter** is a professor in programming languages and programming tools. His current research is mainly situated in the field of distributed programming, concurrent programming, reactive programming and big data processing. His research methodology varies from more theoretical approaches (e.g., type systems) to building practical frameworks and tools (e.g., crowd-sourcing systems). Contact him at Wolfgang.De.Meuter@vub.be.

**Coen De Roover** is an associate professor at the Software Languages Lab of the Vrije Universiteit Brussel. The central theme of his research is the design of program analysis and transformation techniques, and their application in development tools that vary in scope from a single program, over the program's development history, to the evolution of an entire ecosystem of programs over time. He has published over 80 peer-reviewed articles in the domain, and he is actively involved in collaborative research projects of a fundamental, strategic, and applied nature. He frequently serves on the program committee for international conferences such as ASE, MSR, ICSME, SANER, and SCAM. Contact him at Coen.De.Roover@vub.be.